\documentclass[preprint2,times,tighten]{aastex6}
\pdfoutput=1 
\usepackage{amsmath,amstext}
\usepackage{xspace}
\usepackage{amssymb}
\usepackage{graphicx} 
\usepackage[T1]{fontenc}
\usepackage[all]{hypcap} 

\newcommand{\teff}{$T_{\mathrm{eff}}$\xspace} 
\newcommand{\aFe}{[$\alpha$/M]\xspace} 
\newcommand{\al}{$\alpha$} 

\shorttitle{A radial age gradient in the geometrically thick disk of the Milky Way}
\shortauthors{Martig et al.}

\begin{document}

\title{A radial age gradient in the geometrically thick disk of the Milky Way}
\author{Marie Martig\altaffilmark{1}, Ivan Minchev\altaffilmark{2}, Melissa Ness\altaffilmark{1}, Morgan Fouesneau\altaffilmark{1}, Hans-Walter Rix\altaffilmark{1}}
\email{marie.martig@gmail.com}
\altaffiltext{1}{Max-Planck-Institut f\"{u}r Astronomie, K\"{o}nigstuhl 17, D-69117 Heidelberg, Germany}
\altaffiltext{2}{Leibniz-Institut f\"{u}r Astrophysik Potsdam (AIP), An der Sternwarte 16, D-14482 Potsdam, Germany}

\begin{abstract}
In the Milky Way, the thick disk can be defined using individual stellar abundances, kinematics, or age; or geometrically, as stars high above the mid-plane. In nearby galaxies, where only a geometric definition can be used, thick disks appear to have large radial scale-lengths, and their red colors suggest that they are uniformly old. The Milky Way's geometrically thick disk is also radially extended, but it is far from chemically uniform: \al-enhanced  stars are confined within the inner Galaxy. 
In simulated galaxies, where old stars are centrally concentrated, geometrically thick disks are radially extended, too. Younger stellar populations flare in the simulated disks' outer regions, bringing those stars high above the mid-plane.
The resulting geometrically thick disks therefore show a radial age gradient, from old in their central regions to younger  in their outskirts. 
Based on our age estimates for a large sample of giant stars in the APOGEE survey, we can now
test this scenario for the Milky Way.  We  find that the geometrically-defined thick disk in the Milky Way has indeed a strong radial age gradient: the median age for red clump stars goes from $\sim 9$ Gyr in the inner disk to 5 Gyr in the outer disk. We propose that at least some nearby galaxies could also have thick disks that are not uniformly old, and that geometrically thick disks might be complex structures resulting from different formation mechanisms in their inner and outer parts.
\end{abstract}
 
\maketitle
\section{Introduction}
Thick disks have now been known to exist for more than 30 years, both in nearby galaxies \citep{Burstein1979,Tsikoudi1979} and in the Milky Way \citep{Gilmore1983}. However, there are different ways to define a thick disk:
\begin{itemize}
\item geometrically (or morphologically), based on decomposition of vertical density profiles \citep{Gilmore1983, Juric2008, Yoachim2006,Comeron2011}, or at a fixed height above the disk mid-plane \citep{Yoachim2008,Rekjuba2009, Cheng2012a}
\item kinematically \citep{Morrison1990, Majewski1992,Bensby2003, Reddy2003,Adibekyan2012,Haywood2013}
\item chemically, as the \al-rich sequence in the [\al/Fe] vs [Fe/H] plane \citep{Fuhrmann1998,Navarro2011,Adibekyan2012,Bovy2012}
\item as the old part of the disk \citep{Haywood2013,Bensby2014,Xiang2015}  
\end{itemize}

While all of these definitions can be applied in the Milky Way, only a geometric definition can be used for external galaxies. These geometrically thick disks are extended (they form a red envelope all around the thin disks) and have scale-lengths comparable to those of thin disks \citep{Yoachim2006,Pohlen2007,Comeron2012}. Their red colors (and the absence of radial color gradients) have led to the tentative conclusion that they are made of uniformly old stellar populations \citep{Dalcanton2002,Rekjuba2009}. However, the degeneracy between age and metallicity measured from broad-band photometry complicates further exploration of the age and chemical structure of thick disks in nearby galaxies.

In the Milky Way, the geometrically defined thick disk has a large scale-length \citep[$\sim$ 3.5--4 kpc,][]{Ojha2001, Juric2008, Jayaraman2013}, in agreement with measurements for nearby galaxies. 
However, in the Milky Way this geometrically thick disk does not correspond to a uniform physical component in terms of chemical properties. Indeed, the \al-rich thick disk is centrally concentrated with a short scale-length of about 2 kpc \citep{Bensby2011, Cheng2012b, Bovy2012,Bovy2016}, and very few \al-rich stars are found in the outer disk of the Milky Way \citep{Nidever2014, Hayden2015}. This means that the chemically-defined thick disk and the geometrically-defined thick disk have a totally different structure.
While this discrepancy  has been mentionned by several authors \citep[e.g.,][]{Bovy2012,Jayaraman2013}, the reasons for the discrepancy itself have received less attention.

In \cite{Minchev2015} we used numerical simulations to propose an explanation. We showed that in simulated disks the oldest stellar populations are indeed concentrated within the inner disk, while younger stellar populations have larger scale-lengths and smaller scale-heights \citep[see also][]{Martig2014}. However, we also showed that a thin-thick disk decomposition is still possible even in the outer disk, and that such geometrically-defined thick disks are very extended.
This is because most mono-age populations flare in their outer regions, with the flaring radius increasing for younger populations (such a flaring was recently found by \citealp{Bovy2016} for \al-poor stars in the Milky Way). As a consequence, while the very center of the galaxy is dominated by old stars, the more extended parts of the thick disk are made of progressively younger stellar populations, so that a geometrically defined thick disk would have a radial age gradient going from old stars in the center to young stars in the outskirts of the galaxy. Such a radial age gradient has also been seen independently in simulations by \cite{Rahimi2014} and \cite{Miranda2015}. 

However, we lack a direct observational test of this theoretical picture using actual stellar ages instead of abundance proxies. 
Two studies (\citealp{Martig2016}, M16, and \citealp{Ness2016}, N16) have recently (and for the first time) determined ages for stars over a large volume of the Galaxy within the  Apache Point Observatory Galactic Evolution Experiment (APOGEE) survey \citep{Majewski2015}. In this paper we use these two sets of stellar ages to show that in the Milky Way the geometrically-defined thick disk indeed shows a radial age gradient, as predicted by the simulations.

In Section 2, we present the APOGEE data and the techniques used to derive ages. We then present in Section 3 our results on the age structure of the disk of the Milky Way. In Section 4, we discuss the robustness of our results, compare the Milky Way to nearby galaxies and conclude the paper with a discussion of the implications of our results for thick disk formation scenarios.

\section{Data and analysis}

We use a sample of red giants selected from the APOGEE Data Release 12 \citep{Holtzman2015}. APOGEE is a high-resolution ($R=22,500$) spectroscopic survey in the H-band using the 2.5m SDSS telescope. The spectra are treated by the APOGEE Stellar Parameter and Chemical Abundances Pipeline (ASPCAP, \citealp{GarciaPerez2015}),  providing stellar parameters (\teff, $\log g$, [M/H], \aFe, [C/M], and [N/M]), as well as 15 element abundances for over 150,000 stars. In addition to these parameters, we have recently determined ages for $\sim$52,000 of the APOGEE red giants using two independent methods (M16, N16).

Both studies use as a training set a sample of $\sim$1,500 stars from the APOKASC survey \citep{Pinsonneault2014}, which combines spectroscopic information from APOGEE and asteroseismic information from the \textit{Kepler} Asteroseismic Science Consortium (KASC). This unique combination allows for a good determination of stellar masses, and by extension, of stellar ages (using stellar evolution models).

Using the APOKASC sample, M16 determined an empirical relation between the mass (and thus age) of red giants and their surface properties. In M16, we built a model predicting mass and age as a function of [M/H], [C/M], [N/M], [C+N/M], $\log g$ and \teff. From cross-validation, we established that this model predicts masses with an r.m.s error of 12\%  (42\% for ages). 
We then applied this model to 52,286 giants in the rest of APOGEE DR12 for which no seismic data (and hence no precise mass and age information) is available. We restrict ourselves to regions of the parameter space covered by our training set.

By contrast, N16 determined stellar ages directly from the spectra using \textit{The Cannon} \citep{Ness2015}. From the training set, \textit{The Cannon} builds a model that maps stellar parameters to the flux as a function of wavelength. N16 have shown that they can extract age information from the APOGEE spectra with an accuracy of 40\%, similarly to M16.

In this paper, we mostly focus our analysis on a sample of 14,685 red clump (RC) stars for which distances are determined with a precision of  5--10\% by \cite{Bovy2014}. We also use the larger red giant branch (RGB) sample, with distances computed by \cite{Ness2016a} with a precision of $\sim$30\%.

\begin{figure*}
\centering 
\includegraphics[width=0.7\textwidth]{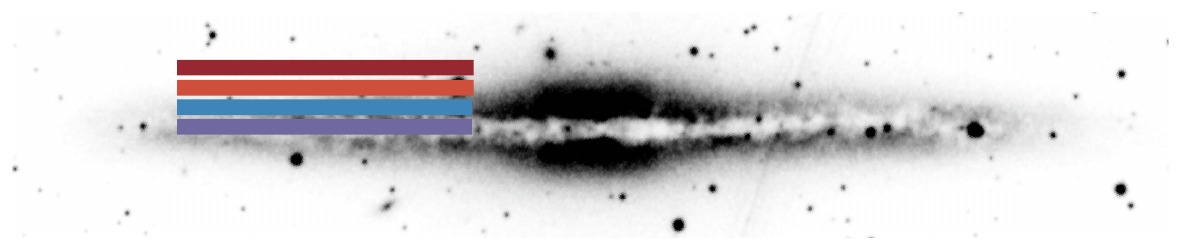}
\includegraphics[width=0.48\textwidth]{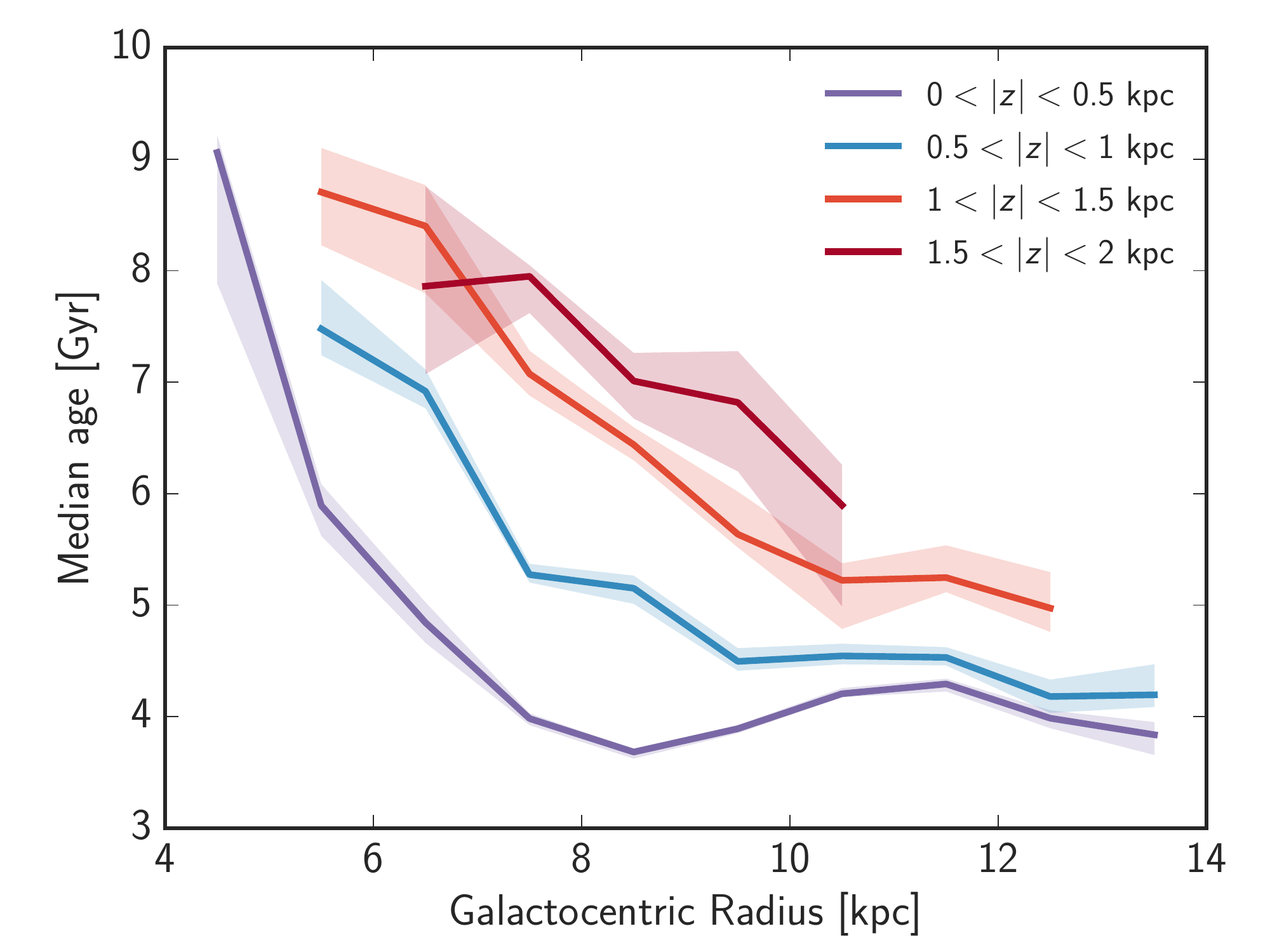}
\includegraphics[width=0.48\textwidth]{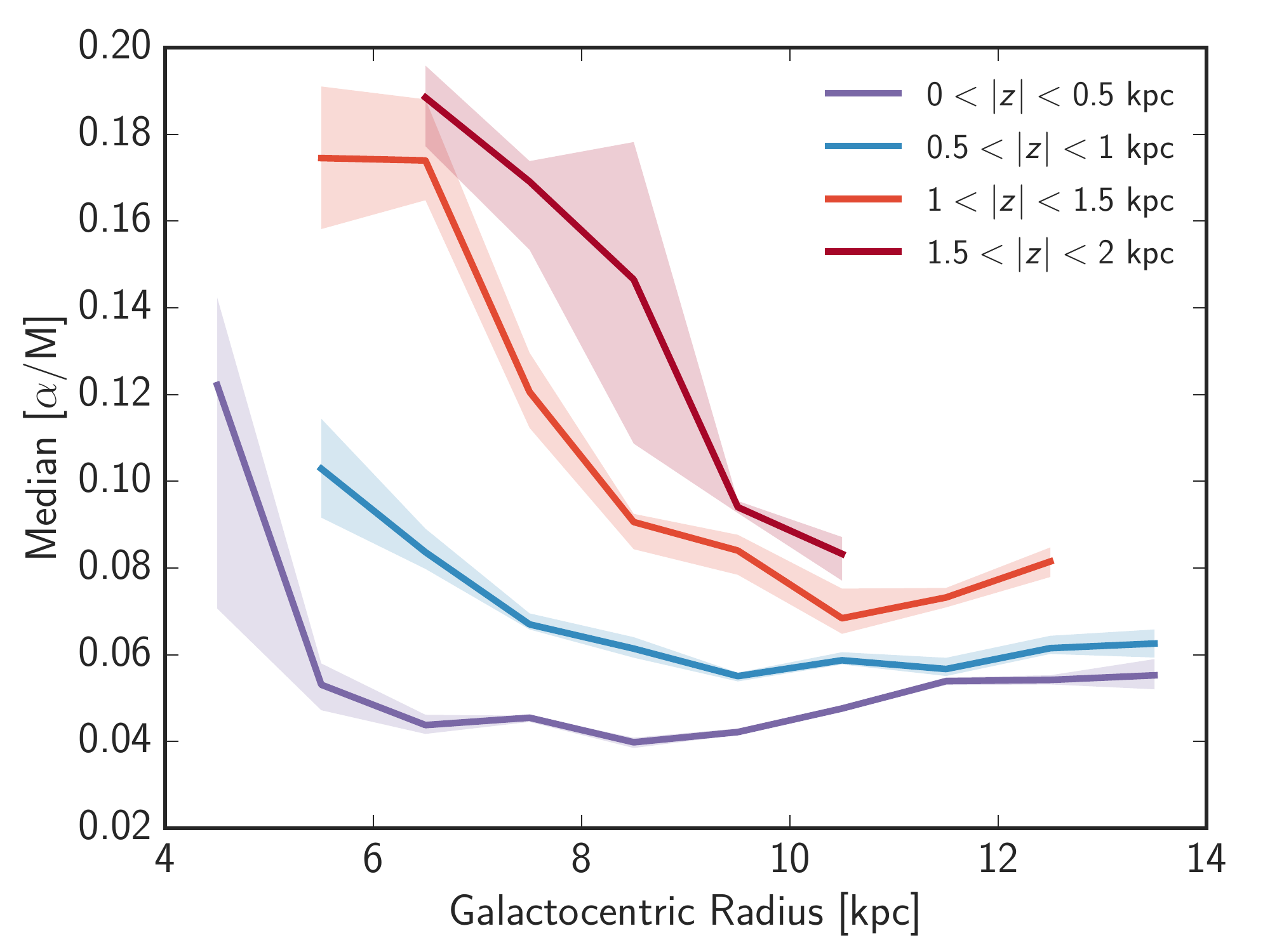}
\caption{Radial profiles of age (left) and \aFe (right) for RC stars at different heights from the mid-plane of the Milky Way in the APOGEE survey. The solid lines correspond to the median values in each bin, while the shaded areas represents the uncertainty on these medians (the range from the 16th to 84th percentiles, based on 1000 bootstrap samples). The top image is a DSS image of NGC891, and illustrates the physical location of our different $z$ slices. At all heights above the disk, we find a radial gradient of age and \aFe.}
\label{fig:gradients_RC}
\end{figure*}
\section{Structure of the geometrically-defined thick disk}

At the solar radius, the mean metallicity of stars decreases as a function of height above the mid-plane, while the mean \aFe increases \citep[e.g.,][]{Gilmore1985,Ivezic2008,Bovy2012,Schlesinger2012}, and the mean age increases \citep{Casagrande2016}.  
The geometrically-defined thick disk (typically, stars further than 1 kpc from the mid-plane) is thus locally made of stars that are metal-poor, $\alpha$-rich and old.

Such a simple picture does not hold over the whole extent of the Milky Way. The chemical abundance structure of the disk of the Milky Way has already been studied extensively, notably by \cite{Nidever2014} using the APOGEE RC sample and by \cite{Hayden2015} using all the DR12 red giants. These studies confirm that at all galactocentric radii metallicity decreases and \aFe increases with height above the mid-plane, but a striking feature is the nearly total disappearance of $\alpha$-rich stars in the outer disk (beyond 11 kpc). This means that the geometrically-defined thick disk, while having overall a flat radial metallicity gradient  \citep[see also][]{Cheng2012a}, has a strong radial \aFe gradient, and is mostly $\alpha$-poor in its outer regions. This is already a clear indication of the complex nature of the thick disk, and we show here for the first time that this complexity is also found in terms of age structure.

We split the 14,685 RC stars into four bins corresponding to distance from the midplane, from 0 to 2 kpc (see the top panel of Figure \ref{fig:gradients_RC} for an indication of the spatial location of the slices superimposed on a DSS image of NGC 891). For each slice, we compute the median age (using the M16 method) and \aFe as a function of radius in 1 kpc-wide radial bins, only showing in Figure \ref{fig:gradients_RC} the bins with more than 20 stars. To estimate the uncertainty on the median in each bin, we draw 1000 bootstrap realizations of the sample, compute the median age or \aFe  for each realization, and then show in Figure \ref{fig:gradients_RC} the range containing the 16$^{\mathrm{th}}$ to 84$^{\mathrm{th}}$ percentiles of all these medians.

We caution that we are here studying the age distribution of RC stars, which differs from the age distribution of the total underlying stellar population (see Figure 15 of \citealp{Bovy2014}). This would need to be taken into account to directly compare our results to simulations. However, in this paper, we just aim at establishing the existence of a radial age gradient in the geometric thick disk, and not at providing accurate absolute ages.

The radial age and \aFe profiles are shown in the bottom panels of Figure \ref{fig:gradients_RC}. At the solar radius, we find a median age for RC stars of $\sim$ 4 Gyr within the mid-plane, increasing to 7.5 Gyr for 1.5$<|z|<$2 kpc. This is roughly consistent with  a vertical age gradient of  $4 \pm 2$ Gyr kpc$^{-1}$ measured for giant stars at the solar radius by \cite{Casagrande2016}.

Outside of the solar neighborhood, a first interesting result is that at any given radius, the median age of RC stars increases with height above the disk.  The vertical age gradients are shallower in the outer disk, where stellar populations look more uniform as a function of height.  We also find radial age gradients at all heights above the mid-plane. At $\sim$1--2 kpc above the disk, stellar ages go from $\sim$8--9 Gyr in the inner disk to $\sim$ 5 Gyr in the outer disk. As already discussed, this radial age gradient is accompanied with a radial \aFe gradient (right panel in Figure \ref{fig:gradients_RC}).

\begin{figure}
\centering 
\includegraphics[width=0.48\textwidth]{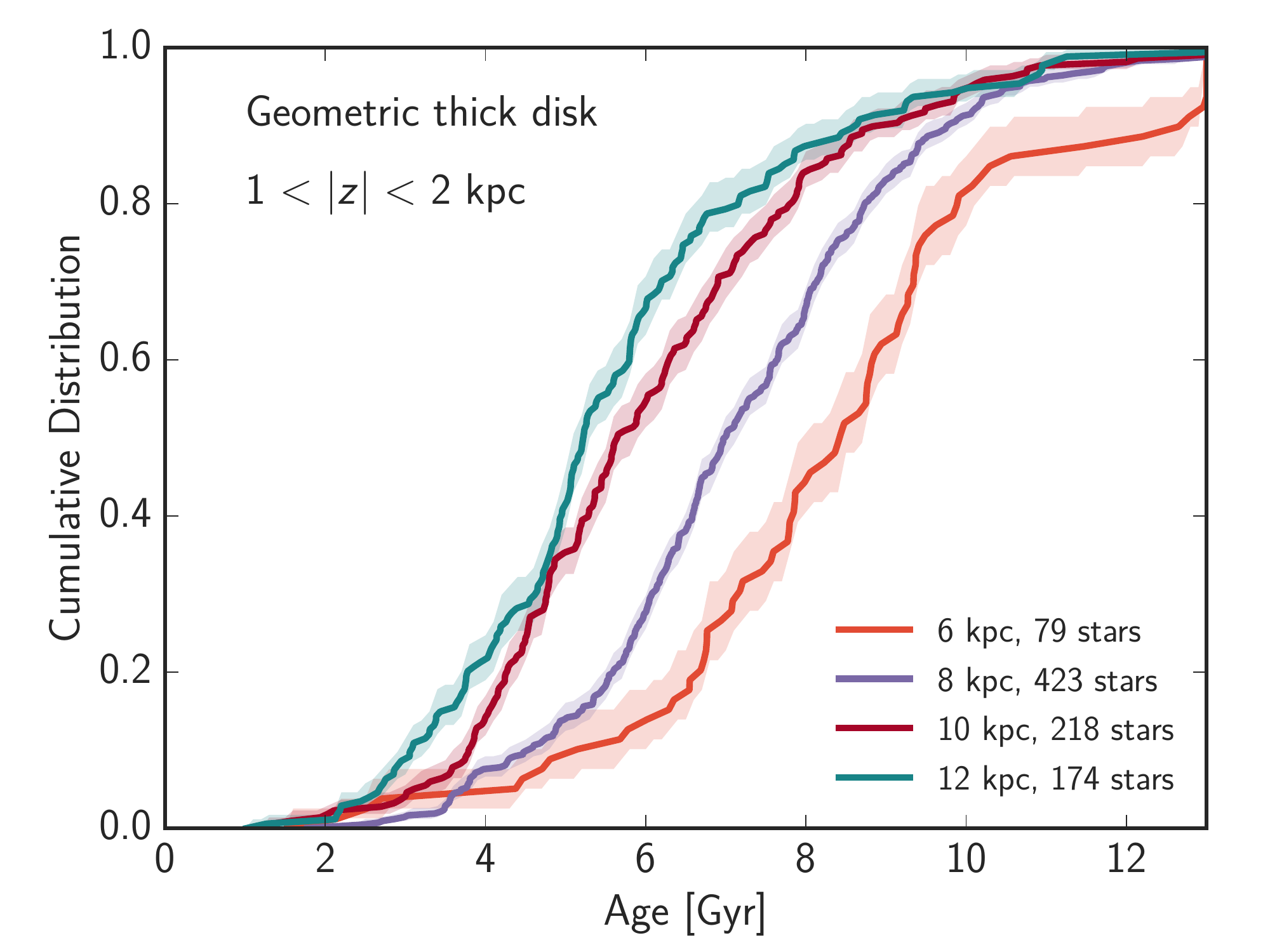}
\includegraphics[width=0.48\textwidth]{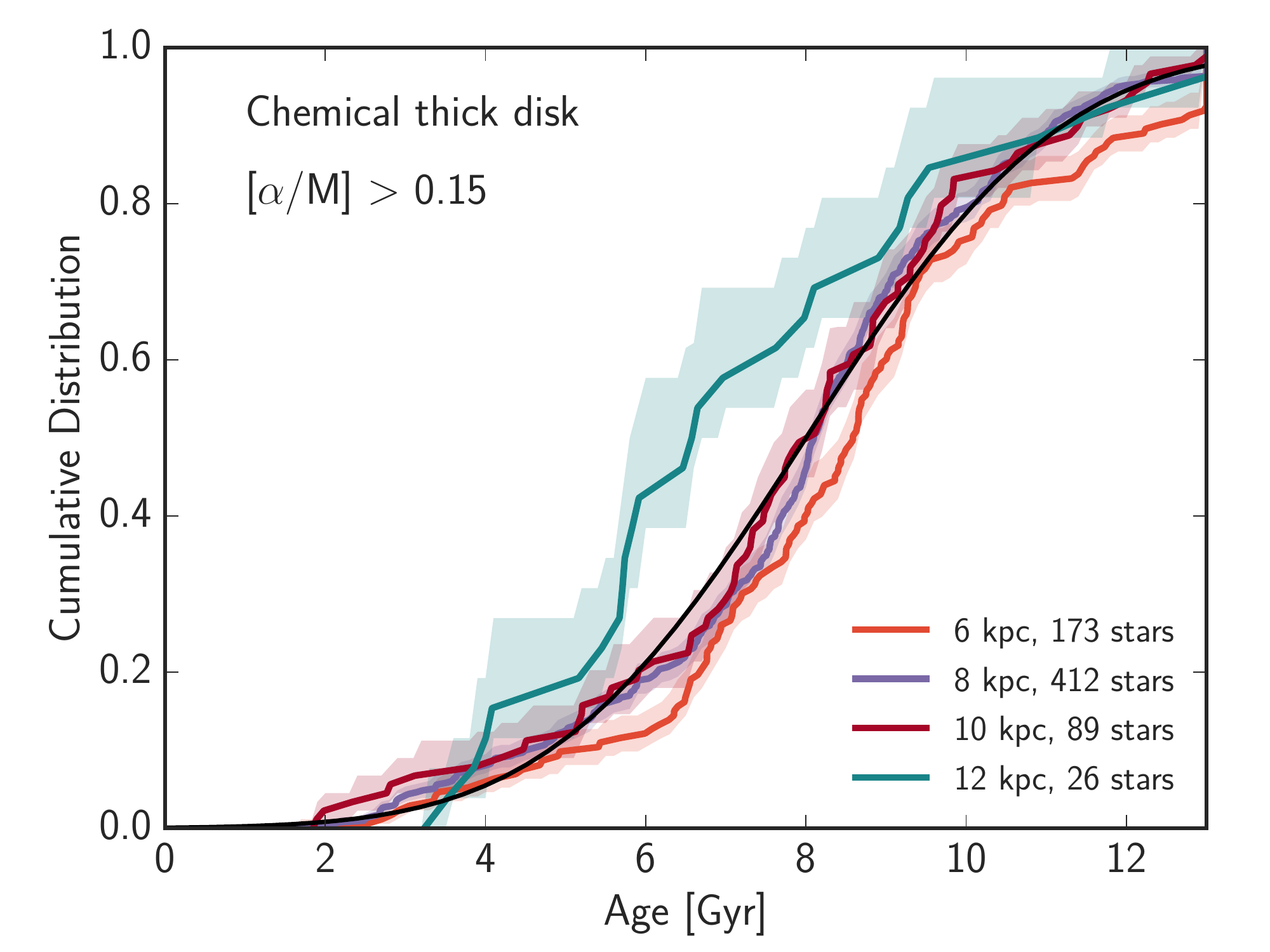}
\caption{Cumulative age distributions for RC stars at different galactocentric radii. The shaded areas represent for each distribution the 1-$\sigma$ range from 1000 bootstrap samples. The top panel shows stars in the geometrically-defined thick disk (i.e., stars far from the mid-plane) while the bottom panel shows stars in the chemically-defined thick disk (i.e., \al-rich stars). The \al-rich population is quite uniform as a function of galactocentric radius, with an age distribution consistent with a 2.5 Gyr-wide Gaussian centered on 8 Gyr (black line). By contrast, the age distribution of the geometrically-defined thick disk changes significantly as a function of radius, with younger stars in the outer regions.}
\label{fig:cumul}
\end{figure}

The top panel in Figure \ref{fig:cumul} shows how the age distribution of stars in the geometric thick disk changes with galactocentric radius. The shaded regions represent the 1-sigma range obtained from 1000 bootstrap realizations of our sample. The age distributions at all radii are significantly different, with younger ages towards the outer disk. The fraction of stars younger than 6 Gyr goes from 15\% in the inner disk to 70\% in the outer disk. These radial age variations are very similar to what we found in our simulations (Fig. 2 in \citealp{Minchev2015}), although a direct comparison needs to take into account the data selection function and the age distribution of RC stars.

The age gradient is not a consequence of a change in the ages of \al-rich stars. The bottom panel in Figure \ref{fig:cumul} shows that the age distribution of \al-rich stars is independent of radius (except maybe  at large radii, but this is based only a very small number of stars). This age distribution is roughly consistent with a Gaussian centered on 8 Gyr, with a standard deviation of 2.5 Gyr (black line in this Figure), which would correspond to a 31\% age error. The chemically-defined thick disk is thus remarkably homogeneous in terms of age and seems to form a uniform population (but see \citealp{Liu2012}, finding two families of stars in terms of orbital eccentricity within the \al-rich population, which suggests that the chemically-defined thick disk might be more complex than suggested here).

We thus find that the geometrically-defined thick disk changes in terms of age as a function of radius, and that this age change can be traced to the radial decrease in the fraction of \al-rich stars away from the disk mid-plane.  Therefore, the geometrically and the age-defined thick disks in the Milky Way have fundamentally different structures.

\section{Discussion}
\subsection{Robustness of our results}
The current implementation of our age determination technique does not allow for a measure of the age uncertainties on a star-by-star basis, which prevents us from performing a proper study of how our results are affected by age uncertainties. However, as described in M16, we used a leave-one-out cross validation algorithm to estimate that the r.m.s age error for our training set is $\sim$40\%. We find that the radial age gradients are still present if we create mock data samples by convolving our ages with a 40\%-wide Gaussian error.

\begin{figure}
\centering 
\includegraphics[width=0.48\textwidth]{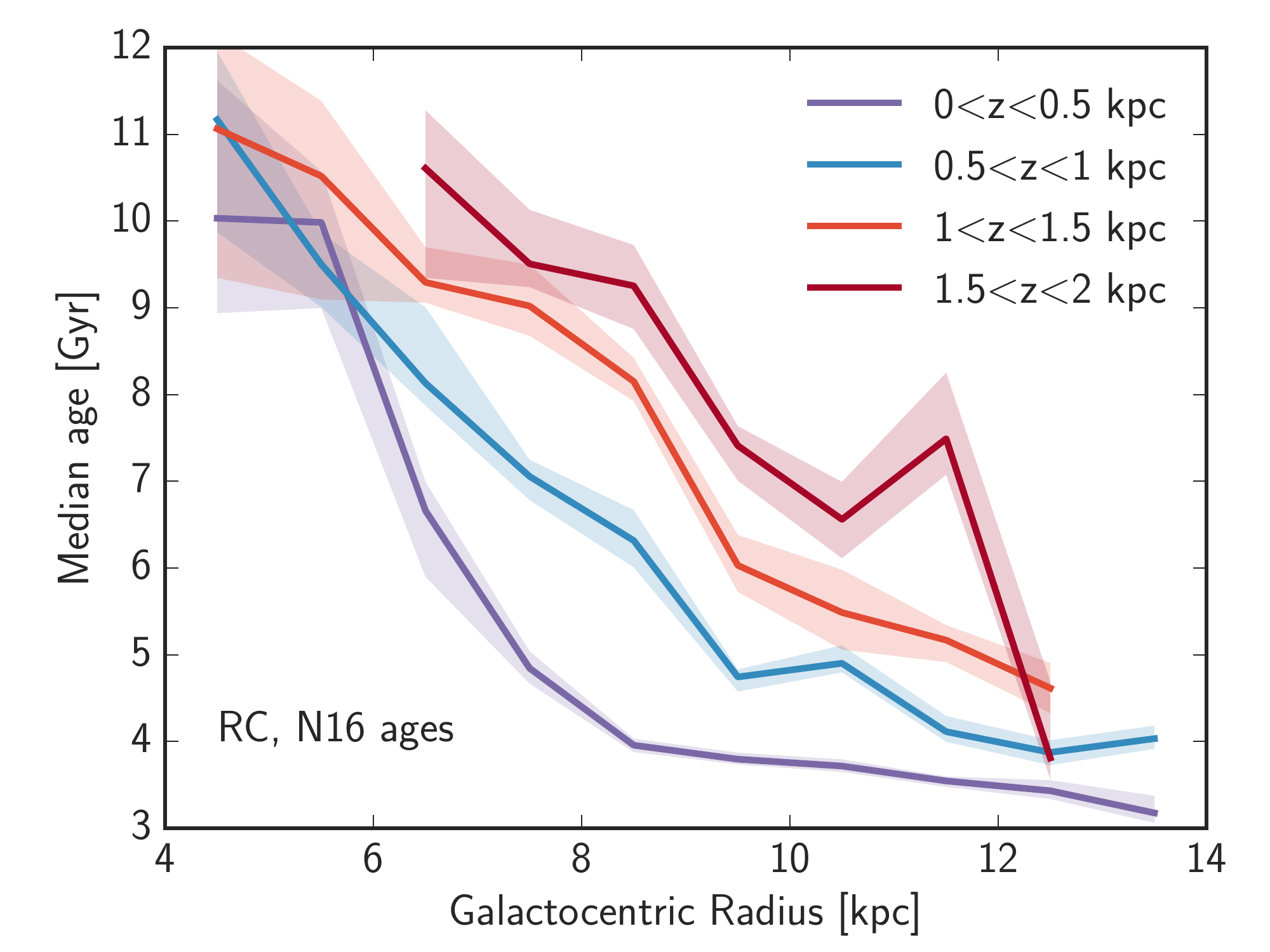}
\includegraphics[width=0.48\textwidth]{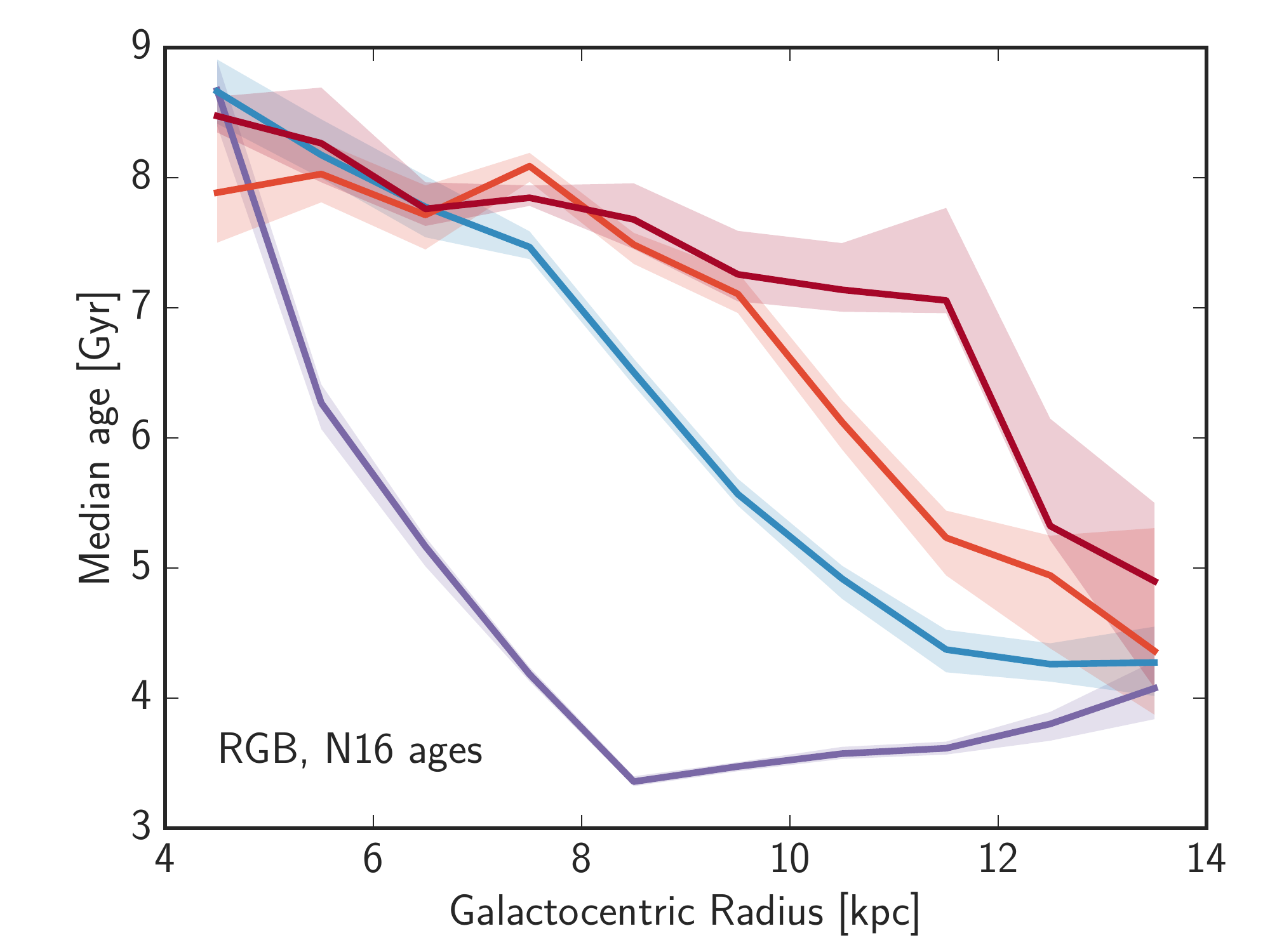}
\caption{Radial age gradients using stellar ages computed by N16 using \textit{The Cannon}, for RC stars on the top panel and for RGB stars on the bottom panel. This confirms the presence of strong radial age gradients at all heights above the mid-plane. The errors on distances are larger for RGB stars, so that the radial gradients are shallower than for RC stars.}
\label{fig:Cannon}
\end{figure}

We also test if our results on the age gradient depend on the method used to determine stellar ages. We repeat our analysis of RC stars using ages obtained by N16 via  \textit{The Cannon}  (see top panel of Figure \ref{fig:Cannon}).  With the N16 ages, the age gradient in the geometrically-defined thick disk is still present --- it is even steeper, with older ages for thick-disk stars in the inner disk. There is, however, a good general agreement between the two age determination techniques, which is reassuring.

Finally, we check that our results are not an artifact related to the use of RC stars. This could arise either from the age values themselves (less robust for RC stars because ages are affected by mass loss during the RGB phase), or from the fact that RC ages are a biased sampling of the underlying total stellar population. We show in the bottom panel of Figure \ref{fig:Cannon} the age gradients for RGB stars (defined as giants in APOGEE DR12 but not in the RC catalogue). We use ages determined by N16 and distances from \cite{Ness2015}. The age gradients are also found for RGB stars, although the gradients are shallower and the shape of the radial trends is slightly different: this reflects the different age distribution of RC vs RGB stars, but also the $\sim$3 times larger distance uncertainties for RGB stars compared to RC stars. 

Using both a different set of stellar ages and a different type of stars, we thus confirm that the geometrically-defined thick disk is younger in its outer regions.
We emphasize again that the median age we find for RGB and RC stars is in no way representative of the age of the underlying total stellar population, and as such cannot be directly compared to simulations. The main obstacle is not so much the survey selection function  (as discussed in \citealp{Hayden2015}, the survey selection function does not depend strongly on metallicity and the sample of giants observed is representative of the underlying population of giants), but rather the complex age distribution of RGB and RC stars. The age distribution of RC and RGB stars tends to be biased towards younger ages, but the strength of the effect depends on the stellar evolutionary phase and the local star formation history \citep{Girardi2001, Bovy2014, Hayden2015}.  Correcting for this age bias would require some complex modelling  which is beyond the scope of this paper. We note however that the age bias does not affect our main result, i.e. the existence of an age difference between the inner and outer disk.

\subsection{The Milky Way compared to nearby galaxies}

Our results show that the geometrically-defined thick disk in the Milky Way has a strong radial age gradient. This reconciles measurements of a short scale-length for the \al-rich disk with measurements of a large scale-length for the geometrically thick disk.  Large scale-lengths are also measured for (geometrically) thick disks in external galaxies, but we do not know yet if these large scale-lengths have the same origin as in the Milky Way (in which case the disks would have a radial age gradient), or if these external geometrically thick disks are uniformly old components. Given the variety of formation histories for disk galaxies \citep[e.g.,][]{Martig2012}, it is likely that both types of thick disks exist. The most direct way to test this would be to identify which nearby galaxies also have an age gradient in their geometrically-defined thick disk. However, measuring ages for thick disks outside of the Milky Way is extremely challenging.

A few Hubble Space Telescope (HST) studies have measured the properties of resolved stars in nearby edge-on disk galaxies, and found older stars at large scale-heights, but do not probe the radial structure of the thick disk \citep{Seth2005, Tikhonov2005, Mould2005}. An exception is \cite{Rekjuba2009}, who study RGB stars in NGC 891 with the HST and do not find any radial color or metallicity gradient along the thick disk.

Spectroscopic studies are limited by the very faint surface brightness of the outer regions of thick disks. The Lick indices study of \cite{Yoachim2008b} was not able to probe the radial age structure of thick disk.  Similarly, while Integral Field Unit spectroscopy is the ideal tool to probe thick disks, the VLT/VIMOS observations of \cite{Comeron2015} were limited by S/N, grouping the entire thick disk region in a single bin, and were thus unable to unveil its structure.
 
\begin{figure}
\centering 
\includegraphics[width=0.48\textwidth]{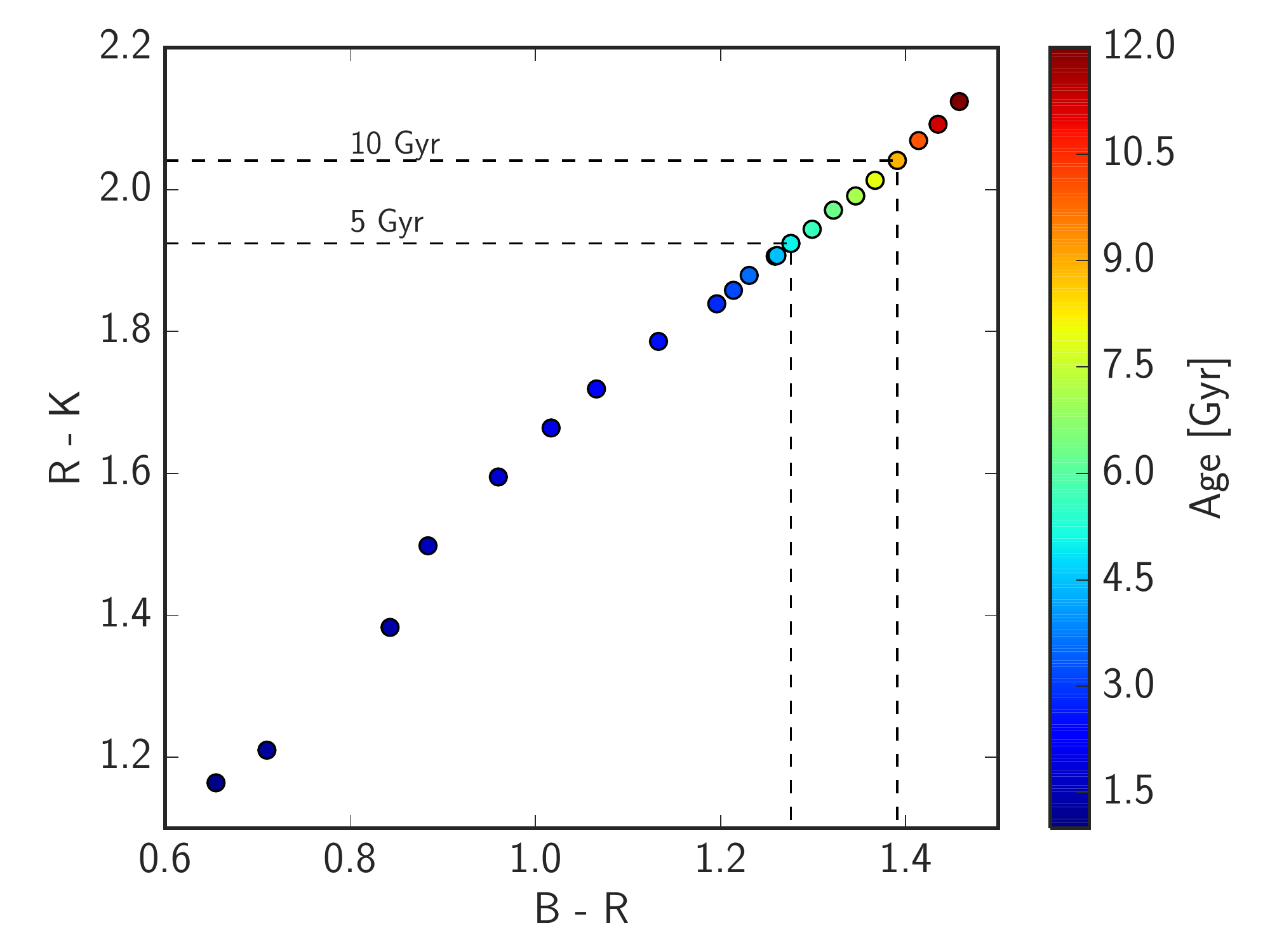}
\caption{Integrated colors as a function of age for a SSP with a metallicity of -0.5 (based on a Chabrier IMF and the PARSEC stellar evolution models). The color difference between populations of 5 and 10 Gyr is of only $\sim$0.1 mag. If nearby edge-on galaxies had the same radial age gradient as the Milky Way, this would not be clearly obvious from their broad-band colors.}
\label{fig:colors}
\end{figure}

Broad-band photometry can more easily reach deeper levels of surface brightness, but age and metallicity are degenerate and age determinations are quite approximate. \cite{Dalcanton2002} measured the B-R and R-K colors of thick disks, finding that thick disks typically have red colors (B-R $\sim$1.3--1.5) and no strong radial  color gradient.
This absence of a color gradient \citep[as also found by][]{Rekjuba2009} could naively be interpreted as an argument against an age gradient. However, these colors are very insensitive to age for populations older than $\sim$5 Gyr. We test this using the PARSEC isochrones \citep{Chen2014} combined with a Chabrier IMF.  As an example, we show in Figure \ref{fig:colors} the variation of B-R and R-K colors for a single stellar population (SSP) of increasing age and a metallicity of -0.5 (roughly typical of a thick disk population). An age gradient from 10 to 5 Gyr along the thick disk would only give a small change in color of $\sim$0.1 mag. 

This means that current broad-band observations cannot exclude younger ages for the outer parts of thick disks and that deeper spectroscopic observations would be needed to probe the age structure of thick disks.

\subsection{Final words: implications for thick disk formation scenarios}

We leave to a future paper the detailed comparison between the age structure of the thick disk in the Milky Way and in simulated galaxies, which  will require a careful modeling of the selection function of our RC sample. However, to first order, the observed age gradient in the geometrically thick disk is qualitatively consistent with  what we find in our simulations \citep{Martig2014,Minchev2015} and also independently in simulations by \cite{Rahimi2014} and \cite{Miranda2015}. This suggests that complex age structures in thick disks might be a common feature of disk galaxy evolution.  We  leave for future work the understanding of the relation between the age structure of a thick disk and its detailed formation history.

A first conclusion we can already draw is that geometrically thick disks (certainly in the Milky Way and maybe  also in some external galaxies) might arise from a succession of events of different nature, and do not need to form all at once at high redshift. The inner parts might have formed in a violent phase at high redshift (either via disk instabilities or mergers), while the outer parts formed later, from the flaring of younger and more extended populations. To test this for external galaxies will require deep spectroscopic observations which measure either age or \aFe profiles along their thick disk (metallicity is not a good indicator of formation history: the Milky Way's geometrically thick disk has a flat metallicity gradient but a complex formation history). This should now be possible with instruments like MUSE on the VLT and will allow for direct tests of the similarity between the Milky Way and its neighbors.

\acknowledgments
We thank the referee for thoughtful comments that have improved the presentation of our results.
The research has received funding from the European Research Council under the European Union's Seventh Framework Programme (FP 7) ERC Grant Agreement n. [321035]. 

Funding for SDSS-III has been provided by the Alfred P. Sloan Foundation, the Participating Institutions, the National Science Foundation, and the U.S. Department of Energy Office of Science. The SDSS-III Web site is http://www.sdss3.org/. SDSS-III is managed by the Astrophysical Research Consortium for the Participating Institutions of the SDSS-III Collaboration including the University of Arizona, the Brazilian Participation Group, Brookhaven National Laboratory, Carnegie Mellon University, University of Florida, the French Participation Group, the German Participation Group, Harvard University, the Instituto de Astrofisica de Canarias, the Michigan State/Notre Dame/JINA Participation Group, Johns Hopkins University, Lawrence Berkeley National Laboratory, Max Planck Institute for Astrophysics, Max Planck Institute for Extraterrestrial Physics, New Mexico State University, New York University, Ohio State University, Pennsylvania State University, University of Portsmouth, Princeton University, the Spanish Participation Group, University of Tokyo, University of Utah, Vanderbilt University, University of Virginia, University of Washington, and Yale University.

{}

\end{document}